# On the equivalence of proportional-integral and proportional-resonant controllers with anti-windup

C.M. Hackl*

**Abstract**

It is shown that proportional-integral (PI) control in the synchronously rotating $(d,q)$-reference frame and proportional-resonant (PR) control in the stationary $(\alpha,\beta)$-reference frame, both with anti-windup, are equivalent *if and only if* their implementation is done correctly in *state space* and the controller parameters and the initial values are identical. It is shown that an equivalence in the frequency domain does only hold if simplifying assumptions are satisfied. As consequence of the equivalence, both closed-loop control performances are *identical* with respect to closed-loop dynamics and steady-state accuracy. The control performance will only differ if their implementation is not done correctly or the time delay induced by the voltage source inverter becomes significant. To the best knowledge of the author, equivalence of PR and PI controllers with anti-windup has not been shown before (in particular not in *state space*).

## Contents

## Notation

$\boldsymbol{x} := (x_1,\dots,x_n)^\top \in \mathbb{R}^n$: column vector, $n \in \mathbb{N}$. $\boldsymbol{0}_n \in \mathbb{R}^n$: zero vector. $\|\boldsymbol{x}\| := \sqrt{\boldsymbol{x}^\top\boldsymbol{x}}$: Euclidean norm of $\boldsymbol{x}$. $\boldsymbol{A} \in \mathbb{R}^{n\times m}$: real matrix, $n,m \in \mathbb{N}$, $\det(\boldsymbol{A})$: determinant of $\boldsymbol{A}$. $\boldsymbol{I}_n \in \mathbb{R}^{n\times n}$: identity matrix. $\mathcal{C}^1(I;Y)$: space of continuously differentiable functions mapping $I \to Y$. For modeling of electrical machines, a signal $\boldsymbol{x}$ may be represented in the three-phase $(a,b,c)$-reference frame $\boldsymbol{x}^{abc} := \left(x^a,\ x^b,\ x^c\right)^\top$, the stationary $(\alpha,\beta)$-reference frame $\boldsymbol{x}^{\alpha\beta} := \left(x^\alpha,\ x^\beta\right)^\top$ and the arbitrarily rotating $(d,q)$-reference frame $\boldsymbol{x}^{dq} := \left(x^d,\ x^q\right)^\top$, which are related by $\boldsymbol{x}^{dq} = \boldsymbol{T}_\mathrm{p}(\phi_\mathrm{k})^{-1}\boldsymbol{x}^{\alpha\beta} = \boldsymbol{T}_\mathrm{p}(\phi_\mathrm{k})^{-1}\boldsymbol{T}_\mathrm{c}\boldsymbol{x}^{abc}$. $\phi_\mathrm{k}$ [rad] is the (electrical) angle of the $k$-reference frame with respect to the $s$-reference frame and $\boldsymbol{T}_\mathrm{p}(\phi_\mathrm{k}) = \begin{bmatrix}\cos(\phi_\mathrm{k}) & -\sin(\phi_\mathrm{k})\\ \sin(\phi_\mathrm{k}) & \cos(\phi_\mathrm{k})\end{bmatrix}$, $\boldsymbol{J} = \begin{bmatrix}0 & -1\\ 1 & 0\end{bmatrix}$ and $\boldsymbol{T}_\mathrm{c} = \frac{2}{3}\begin{bmatrix}1 & -\frac{1}{2} & -\frac{1}{2}\\ 0 & \frac{\sqrt{3}}{2} & -\frac{\sqrt{3}}{2}\end{bmatrix}$ are Park, rotation (by $\frac{\pi}{2}$) and (amplitude correct) Clarke transformation matrix, respectively (see [1, 2]). $x(t) \circ\!\!-\!\!\bullet\, x(s)$ relates a time-varying signal $x(t)$ to its Laplace transform $x(s) := \int_0^\infty x(t)\exp(-s\,t)\,\mathrm{d}t$ (assuming the Laplace transform exists; for more details see [3, Sec. A.3.2]).

*C.M. Hackl is head of the research group "Control of renewable energy systems" (CRES) at the Munich School of Engineering (MSE), Technische Universität München (TUM), Germany. For more information see www.cres.mse.tum.de.

## I. Motivation, problem statement and internal model principle

In [1, App. C.3], the equivalence of proportional-integral (PI) and proportional-resonant (PR) controller is discussed in the frequency domain (including positive and negative sequence). The angular frequency is assumed to be constant. However, this assumption does not hold in general; e.g. for electrical machines or for weak grids (with frequency fluctuations), the electrical angular velocity will rather be time-varying and, hence, *not* constant. In this brief note, the equivalence of PI and PR controller is shown in state space without the need of imposing any assumptions on the angular frequency. Hence, the state space implementations of PI and PR controller can be utilized for machine-side control with rapidly changing electrical angular velocity or for grid-side control of weak grids with time-varying grid frequency.

The following two current control problems are considered: (i) current control of an electrical drive consisting of permanent-magnet synchronous machine (PMSM; possibly with anisotropy) and voltage source inverter (VSI) and (ii) current control of a grid-tied voltage source inverter with RL-filter.

The dynamic models of both problems are well known (neglecting the VSI dynamics):

(i) The dynamic model of a PMSM in the synchronously rotating $(d,q)$-reference frame with permanent-magnet flux linkage orientation is given by (see [4, Chapter 8] or [2] with the same notation as in this paper)

$$\left.\begin{aligned}
\overbrace{\begin{pmatrix} u_\mathrm{s}^d(t)\\ u_\mathrm{s}^q(t)\end{pmatrix}}^{=:\boldsymbol{u}_\mathrm{s}^{dq}(t)} &= R_\mathrm{s}\overbrace{\begin{pmatrix} i_\mathrm{s}^d(t)\\ i_\mathrm{s}^q(t)\end{pmatrix}}^{=:\boldsymbol{i}_\mathrm{s}^{dq}(t)} + \omega_\mathrm{k}(t)\overbrace{\begin{bmatrix}0 & -1\\ 1 & 0\end{bmatrix}}^{=:\boldsymbol{J}}\overbrace{\begin{pmatrix}\psi_\mathrm{s}^d\big(\boldsymbol{i}_\mathrm{s}^{dq}(t)\big)\\ \psi_\mathrm{s}^q\big(\boldsymbol{i}_\mathrm{s}^{dq}(t)\big)\end{pmatrix}}^{=:\boldsymbol{\psi}_\mathrm{s}^{dq}\left(\boldsymbol{i}_\mathrm{s}^{dq}(t)\right)} + \tfrac{\mathrm{d}}{\mathrm{d}t}\boldsymbol{\psi}_\mathrm{s}^{dq}\big(\boldsymbol{i}_\mathrm{s}^{dq}(t)\big), \qquad && \boldsymbol{\psi}_\mathrm{s}^{dq}(\boldsymbol{i}_\mathrm{s}^{dq}(0)) = \boldsymbol{\psi}_\mathrm{s}^{k,0}\in\mathbb{R}^2\\
\tfrac{\mathrm{d}}{\mathrm{d}t}\omega_\mathrm{k}(t) &= \tfrac{n_\mathrm{p}}{\Theta}\Big(m_\mathrm{m}\big(\boldsymbol{i}_\mathrm{s}^{dq}(t)\big) - m_\mathrm{l}(t)\Big), && \omega_\mathrm{k}(0) = n_\mathrm{p}\omega_\mathrm{m}^0\in\mathbb{R}\\
\tfrac{\mathrm{d}}{\mathrm{d}t}\phi_\mathrm{k}(t) &= \omega_\mathrm{k}(t), && \phi_\mathrm{k}(0) = n_\mathrm{p}\phi_\mathrm{m}^0\in\mathbb{R}.
\end{aligned}\right\} \tag{1}$$

with affine flux linkage (in Wb)

$$\boldsymbol{\psi}_\mathrm{s}^{dq}\big(\boldsymbol{i}_\mathrm{s}^{dq}(t)\big) = \underbrace{\begin{bmatrix}L_\mathrm{s}^d & 0\\ 0 & L_\mathrm{s}^q\end{bmatrix}}_{=:\boldsymbol{L}_\mathrm{s}^{dq}\in\mathbb{R}^{2\times 2}}\boldsymbol{i}_\mathrm{s}^{dq}(t) + \underbrace{\begin{pmatrix}\psi_\mathrm{pm}\\ 0\end{pmatrix}}_{=:\boldsymbol{\psi}_\mathrm{pm}^{dq}} \tag{2}$$

and machine torque (in $\mathrm{N\,m}$)

$$m_\mathrm{m}\big(\boldsymbol{i}_\mathrm{s}^{dq}(t)\big) = \tfrac{3}{2}n_\mathrm{p}\,\boldsymbol{i}_\mathrm{s}^{dq}(t)^\top\boldsymbol{J}\boldsymbol{\psi}_\mathrm{s}^{dq}\big(\boldsymbol{i}_\mathrm{s}^{dq}(t)\big) \overset{(2)}{=} \tfrac{3}{2}n_\mathrm{p}\left[\boldsymbol{i}_\mathrm{s}^{dq}(t)^\top\boldsymbol{J}\boldsymbol{L}_\mathrm{s}^{dq}\boldsymbol{i}_\mathrm{s}^{dq}(t) + \boldsymbol{i}_\mathrm{s}^{dq}(t)^\top\boldsymbol{J}\boldsymbol{\psi}_\mathrm{pm}^{dq}\right]. \tag{3}$$

In (1), (2) and (3), $R_\mathrm{s}$ (in $\Omega$) is the stator resistance, $\boldsymbol{u}_\mathrm{s}^{dq} := (u_\mathrm{s}^d, u_\mathrm{s}^q)^\top$ (in V), $\boldsymbol{i}_\mathrm{s}^{dq} := (i_\mathrm{s}^d,\, i_\mathrm{s}^q)^\top$ (in A) and $\boldsymbol{\psi}_\mathrm{s}^{dq} := (\psi_\mathrm{s}^d,\, \psi_\mathrm{s}^q)^\top$ (in Wb) are stator voltage (e.g. applied by a machine-tied voltage source inverter), current and flux linkage vectors, respectively. Note that $\omega_\mathrm{k} = n_\mathrm{p}\,\omega_\mathrm{m}$ (in $\mathrm{rad/s}$) and $\phi_\mathrm{k} = n_\mathrm{p}\phi_\mathrm{m}$ are *electrical* angular frequency and angle, whereas $\omega_\mathrm{m}$ and $\phi_\mathrm{m}$ are *mechanical* angular frequency and angle of the rotor (with initial values $\omega_\mathrm{m}^0$ and $\phi_\mathrm{m}^0$), respectively. $n_\mathrm{p}$ is the pole pair number of the machine and $\Theta$ (in $\mathrm{kg\,m^2}$) is the (rotor's) inertia. $m_\mathrm{m}$ is the electro-magnetic machine torque[1] and $m_\mathrm{l}$ (in $\mathrm{N\,m}$) is a (bounded) load torque. The flux linkage $\boldsymbol{\psi}_\mathrm{s}^{dq}$ depends on the *symmetric*, *positive-definite* inductance matrix $\boldsymbol{L}_\mathrm{s}^{dq} = (\boldsymbol{L}_\mathrm{s}^{dq})^\top > 0$ [6] with stator inductances $L_\mathrm{s}^q > 0$, $L_\mathrm{s}^d > 0$ (both in H), the stator currents $\boldsymbol{i}_\mathrm{s}^{dq}$ and the permanent-magnet flux linkage $\boldsymbol{\psi}_\mathrm{pm}^{dq} = (\psi_\mathrm{pm},\, 0)^\top$. To obtain the current dynamics of the PMSM in the stationary $(\alpha,\,\beta)$-reference frame, the Park transformation $\boldsymbol{T}_\mathrm{p}(\phi_\mathrm{k}(t))$ must be applied to (1) which yields

$$\overbrace{\begin{pmatrix}u_\mathrm{s}^\alpha(t)\\ u_\mathrm{s}^\beta(t)\end{pmatrix}}^{=:\boldsymbol{u}_\mathrm{s}^{\alpha\beta}(t)} = R_\mathrm{s}\overbrace{\begin{pmatrix}i_\mathrm{s}^\alpha(t)\\ i_\mathrm{s}^\beta(t)\end{pmatrix}}^{=:\boldsymbol{i}_\mathrm{s}^{\alpha\beta}(t)} + \overbrace{\boldsymbol{T}_\mathrm{p}\big(\phi_\mathrm{k}(t)\big)\boldsymbol{L}_\mathrm{s}^{dq}\boldsymbol{T}_\mathrm{p}\big(\phi_\mathrm{k}(t)\big)^{-1}}^{=:\boldsymbol{L}_\mathrm{s}^{\alpha\beta}(\phi_\mathrm{k}(t))}\tfrac{\mathrm{d}}{\mathrm{d}t}\boldsymbol{i}_\mathrm{s}^{\alpha\beta}(t) + \overbrace{\psi_\mathrm{pm}\begin{pmatrix}\cos\big(\phi_\mathrm{k}(t)\big)\\ \sin\big(\phi_\mathrm{k}(t)\big)\end{pmatrix}}^{=:\boldsymbol{\psi}_\mathrm{pm}^{\alpha\beta}(t)}, \quad \boldsymbol{i}_\mathrm{s}^{\alpha\beta}(0) = \boldsymbol{i}_\mathrm{s}^{\alpha\beta,0}\in\mathbb{R}^2. \tag{4}$$

(ii) The dynamic model of an RL-filter connected to a balanced grid is given in the synchronously rotating $(d,q)$-reference frame with grid voltage orientation by (neglecting the power flow over the DC-link, see [1, Sec. 9.2.1] or [2] with the same notation as in this paper)

$$\overbrace{\begin{pmatrix}u_\mathrm{f}^d(t)\\ u_\mathrm{f}^q(t)\end{pmatrix}}^{=:\boldsymbol{u}_\mathrm{f}^k(t)} = R_\mathrm{f}\overbrace{\begin{pmatrix}i_\mathrm{f}^d(t)\\ i_\mathrm{f}^q(t)\end{pmatrix}}^{=:\boldsymbol{i}_\mathrm{f}^k(t)} + \omega_\mathrm{k}(t)L_\mathrm{f}\boldsymbol{J}\boldsymbol{i}_\mathrm{f}^k(t) + L_\mathrm{f}\tfrac{\mathrm{d}}{\mathrm{d}t}\boldsymbol{i}_\mathrm{f}^k(t) + \overbrace{\begin{pmatrix}\hat{u}_\mathrm{g}(t)\\ 0\end{pmatrix}}^{=:\boldsymbol{u}_\mathrm{g}^k(t)}, \qquad \boldsymbol{i}_\mathrm{f}^k(0) = \boldsymbol{i}_\mathrm{f}^{k,0}\in\mathbb{R}^2, \tag{5}$$

where $R_\mathrm{f}$ (in $\Omega$) and $L_\mathrm{f}$ (in H) are filter resistance and inductance, respectively; $\boldsymbol{u}_\mathrm{f}^k := (u_\mathrm{f}^d, u_\mathrm{f}^q)^\top$ (in V), $\boldsymbol{i}_\mathrm{f}^k := (i_\mathrm{f}^d,\, i_\mathrm{f}^q)^\top$ (in A) and $\boldsymbol{u}_\mathrm{g}^k := (\hat{u}_\mathrm{g},\, 0)^\top$ (in Wb) are filter voltage (e.g. applied by a grid-tied voltage source inverter), filter current and grid voltage vectors, respectively. Note that $\hat{u}_\mathrm{g}$ (in V) and $\omega_\mathrm{k} = 2\pi f_\mathrm{g}$ (in $\frac{\mathrm{rad}}{\mathrm{s}}$ with grid frequency $f_\mathrm{g}$ in Hz) are grid voltage magnitude and grid angular frequency (both obtained from a phase-locked loop), respectively.
The dynamics in the stationary $(\alpha,\, \beta)$-reference frame are obtained by applying the Park transformation $\boldsymbol{T}_\mathrm{p}(\phi_\mathrm{k}(t))$ with $\phi_\mathrm{k}(t) = \int_0^t \omega_\mathrm{k}(\tau)\,\mathrm{d}\tau$ to (5) and are given by

$$\overbrace{\begin{pmatrix} u_\mathrm{f}^\alpha(t) \\ u_\mathrm{f}^\beta(t) \end{pmatrix}}^{=:\boldsymbol{u}_\mathrm{f}^s(t)} = R_\mathrm{f} \overbrace{\begin{pmatrix} i_\mathrm{f}^\alpha(t) \\ i_\mathrm{f}^\beta(t) \end{pmatrix}}^{=:\boldsymbol{i}_\mathrm{f}^s(t)} + L_\mathrm{f} \tfrac{\mathrm{d}}{\mathrm{d}t} \boldsymbol{i}_\mathrm{f}^s(t) + \overbrace{\hat{u}_\mathrm{g}(t) \begin{pmatrix} \cos\big(\phi_\mathrm{k}(t)\big) \\ \sin\big(\phi_\mathrm{k}(t)\big) \end{pmatrix}}^{=:\boldsymbol{u}_\mathrm{g}^s(t)}, \qquad \boldsymbol{i}_\mathrm{f}^s(0) = \boldsymbol{i}_\mathrm{f}^{s,0} \in \mathbb{R}^2 \tag{6}$$

[1] The factor $3/2$ is due to an amplitude-correct Clarke transformation [5, Sec. 16.7].

For both current control problems, in the $(d, q)$-reference frame or in the $(\alpha, \beta)$-reference frame, either *constant* or *sinusoidal* signals must be tracked, respectively. The "Internal Model Principle"–introduced by W.M. Wonham in the 1970s–postulates that *"every good regulator must incorporate a model of the outside world"* being capable to reduplicate *"the dynamic structure of the exogenous signals which the regulator is required to process"* [7, p. 210]. Examples of such exogenous (external and time-varying) signals might be a constant $c > 0$ with corresponding Laplace transform $c \circ\!\!-\!\!\bullet \frac{c}{s}$ and/or sinusoids $\sin(\omega_\mathrm{k} t)$ with corresponding Laplace transform $\sin(\omega_\mathrm{k} t) \circ\!\!-\!\!\bullet \frac{\omega_\mathrm{k}}{s^2+\omega_\mathrm{k}^2}$ (or $\cos(\omega_\mathrm{k} t) \circ\!\!-\!\!\bullet \frac{\omega_\mathrm{k} s}{s^2+\omega_\mathrm{k}^2}$) [3, Tab. A.3.2]. Clearly, the problems described above already motivate for the use of integral control action found in PI controllers and resonant control action found in PR controllers to compensate for constant and sinusoidal signals in the $(d, q)$-reference frame and the $(\alpha, \beta)$-reference frame, respectively.

Both, PI and PR control, will be discussed in more detail in the following sections. Note that, in the remainder, the subscripts $_\mathrm{s}$ and $_\mathrm{f}$ for stator and filter will be dropped. It will be shown that controller structure and design are very similar for the machine-side and grid-side control problem.

## II. Proportional-integral (PI) controller with anti-windup in the $(d, q)$-reference frame

The well-known proportional-integral (PI) controller with anti-windup is re-visited.

### A. Controller structure (see Fig. 1a)

The PI controller structure (following the idea in [5, Sec. 7.1.1] or [2] with the same notation as here) consists of two parts, i.e.

$$\boldsymbol{u}_\mathrm{ref}^{dq}(t) = \underbrace{\boldsymbol{u}_\mathrm{pi}^{dq}(t)}_{\text{PI controller output}} + \underbrace{\boldsymbol{u}_\mathrm{comp}^{dq}(t)}_{\text{disturbance compensation}}. \tag{7}$$

Hence, the voltage reference $\boldsymbol{u}_\mathrm{ref}^{dq} = (u_\mathrm{ref}^d,\, u_\mathrm{ref}^q)^\top$ – the control input to the VSI – is the sum of the disturbance compensation $\boldsymbol{u}_\mathrm{comp}^{dq} = (u_\mathrm{comp}^d,\, u_\mathrm{comp}^q)^\top$ and the output $\boldsymbol{u}_\mathrm{pi}^{dq} = (u_\mathrm{pi}^d,\, u_\mathrm{pi}^q)^\top$ of the PI controller(s). Both parts will be discussed in more detail in the following.

### B. Disturbance compensation (feedforward control)

The goal of the disturbance compensation is to obtain (almost) *decoupled* current dynamics for controller design in the $(d, q)$-reference frame. Therefore, depending on the application (see Fig. 1a), the coupling or disturbance term [2]

$$\boldsymbol{u}_\mathrm{dist}^{dq}(t) := \begin{cases} -\omega_\mathrm{k}(t) \boldsymbol{J} \big(\boldsymbol{L}_\mathrm{s}^{dq} \boldsymbol{i}_\mathrm{s}^{dq} + \boldsymbol{\psi}_\mathrm{pm}^{dq}\big), & \text{for PMSMs as in (1)} \\ -\omega_\mathrm{g}(t) L_\mathrm{f} \boldsymbol{J} \boldsymbol{i}_\mathrm{f}^k(t) - \boldsymbol{u}_\mathrm{g}^k(t), & \text{for RL-filter \& grid as in (5)} \end{cases} \tag{8}$$

depends on (i) (possibly time-varying) grid voltage vector $\boldsymbol{u}_\mathrm{g}^k(t) = (\hat{u}_\mathrm{g}(t),\, 0)^\top$ and grid angular velocity $\omega_\mathrm{k}(t)$, and filter inductance $L_\mathrm{f}$ for grid-side control or (ii) electrical angular velocity $\omega_\mathrm{k}(t) = n_\mathrm{p}\omega_\mathrm{m}(t)$, stator inductances $L_\mathrm{s}^d$, $L_\mathrm{s}^q$ and permanent-magnet flux linkage $\boldsymbol{\psi}_\mathrm{pm}^{dq} = (\psi_\mathrm{pm},\, 0)^\top$ for machine-side control. The disturbance (8) can be (roughly) compensated for by introducing the following feedforward control action

$$\boldsymbol{u}_\mathrm{comp}^{dq}(t) := \begin{pmatrix} u_\mathrm{comp}^d(t) \\ u_\mathrm{comp}^q(t) \end{pmatrix} = -\boldsymbol{u}_\mathrm{dist}^{dq}(t + T_\mathrm{delay}) \;\circ\!\!-\!\!\bullet\; \boldsymbol{u}_\mathrm{comp}^{dq}(s) = \mathrm{e}^{sT_\mathrm{delay}} \boldsymbol{u}_\mathrm{dist}^{dq}(s) \approx \underbrace{\tfrac{c_0(1+sT_\mathrm{delay})}{1+sT_0}}_{=:F_\mathrm{comp}(s)} \boldsymbol{u}_\mathrm{dist}^{dq}(s), \tag{9}$$

where $0 < c_0 \le 1$ (a tuning parameter to avoid over-compensation) and $0 < T_0 \ll T_\mathrm{delay}$ (a tuning parameter to obtain a causal transfer function $F_\mathrm{comp}(s)$). The delay due to $T_\mathrm{delay} \in \big[\frac{1}{2f_\mathrm{sw}},\, \frac{3}{2f_\mathrm{sw}},\big]$ (in s) is induced by the voltage source inverter (VSI) dynamics and is inversely proportional to the switching frequency $f_\mathrm{sw}$ (in Hz) of the inverter [8, 9].

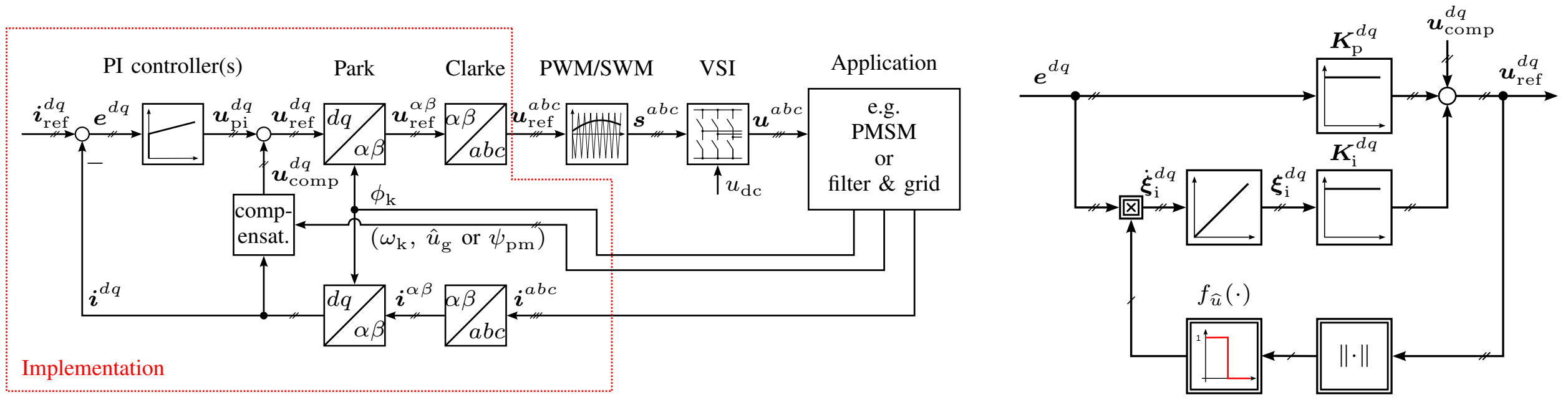

(a) *Overall controller structure.* (b) *PI controller with anti-windup.*

Figure 1: *Block diagrams of the overall controller structure and the PI controller with anti-windup in the synchronously rotating $(d,q)$-reference frame and with disturbance feedforward compensation.*

### *C. PI controller with anti-windup (see Fig. 1b)*

It is well known that PI(D) controllers in presence of input saturation may exhibit integral windup (in particular for large initial errors) leading to large overshoots and/or oscillations in the closed-loop system response (see, e.g., [10, 11]). Due to the limited DC-link voltage $u_{\mathrm{dc}}$ (in V), the output of the VSI is constrained by the saturation level $\widehat{u} \in \left[\frac{u_{\mathrm{dc}}}{2}, \frac{2u_{\mathrm{dc}}}{3}\right]$ (in V) which depends on the employed modulation strategy (such as pulse-width modulation (PWM) or space-vector modulation (SVM) with or without over-modulation [12, Sec. 8.4]).

Due to the input saturation, a simple but effective anti-windup strategy (similar to *conditional integration*, see e.g. [11]) is implemented which stops integration of the integral control action if the control input (here $\boldsymbol{u}^{dq}$ or $\boldsymbol{u}^{dq}_{\mathrm{ref}}$) exceeds the admissible range. For this, the "anti-windup decision function"

$$f_{\widehat{u}} \colon \mathbb{R}_{\geq 0} \to \{0,1\}, \qquad f_{\widehat{u}}\big(\|\boldsymbol{u}^{dq}_{\mathrm{ref}}\|\big) := \begin{cases} 0, & \|\boldsymbol{u}^{dq}_{\mathrm{ref}}\| \geq \widehat{u}, \\ 1, & \|\boldsymbol{u}^{dq}_{\mathrm{ref}}\| < \widehat{u} \end{cases} \tag{10}$$

is combined with the PI controller as follows

$$\left.\boxed{\begin{aligned} \tfrac{\mathrm{d}}{\mathrm{d}t}\boldsymbol{\xi}^{dq}_{\mathrm{i}}(t) &= f_{\widehat{u}}\big(\|\boldsymbol{u}^{dq}_{\mathrm{ref}}(t)\|\big)\, \boldsymbol{e}^{dq}(t), \qquad \boldsymbol{\xi}^{dq}_{\mathrm{i}}(0) = \boldsymbol{\xi}^{dq,0}_{\mathrm{i}} \in \mathbb{R}^2 \\ \boldsymbol{u}^{dq}_{\mathrm{pi}}(t) &= \boldsymbol{K}^{dq}_{\mathrm{p}}\, \boldsymbol{e}^{dq}(t) + \boldsymbol{K}^{dq}_{\mathrm{i}}\, \boldsymbol{\xi}^{dq}_{\mathrm{i}}(t) \end{aligned}}\;\right\} \tag{11}$$

where $\boldsymbol{\xi}^{dq}_{\mathrm{i}} = (\xi^d_{\mathrm{i}}, \xi^q_{\mathrm{i}})^\top$ is the integrator output vector of the PI controller, $\boldsymbol{\xi}^{dq,0}_{\mathrm{i}}$ is its initial value and $\boldsymbol{e}^{dq} = (e^d, e^q)^\top = \boldsymbol{i}^{dq}_{\mathrm{ref}} - \boldsymbol{i}^{dq}$ is the current tracking error. A block diagram of the PI controller (11) with anti-windup is depicted in Fig. 1b. The controller gains are merged into the following diagonal gain matrices

$$\boldsymbol{K}^{dq}_{\mathrm{p}} := \begin{bmatrix} k^d_{\mathrm{p}} & 0 \\ 0 & k^q_{\mathrm{p}} \end{bmatrix} \in \mathbb{R}^{2\times 2} \qquad \text{and} \qquad \boldsymbol{K}^{dq}_{\mathrm{i}} := \begin{bmatrix} k^d_{\mathrm{i}} & 0 \\ 0 & k^q_{\mathrm{i}} \end{bmatrix} \in \mathbb{R}^{2\times 2}. \tag{12}$$

The tuning of the controller gains can be done e.g. according to the "Magnitude Optimum criterion" (see [13] or [5, p. 81,82]) or any other convenient/preferred tuning rule.

**Remark II.1.** *Note that the proportional and integrator gains are* not *necessarily equal, i.e.* $k^d_{\mathrm{p}} \neq k^q_{\mathrm{p}}$ *and* $k^d_{\mathrm{i}} \neq k^q_{\mathrm{i}}$*. In [14] and [15], it was shown that a different choice of the proportional controller gains is beneficial in order to obtain an improved control performance (in particular for anisotropic and/or nonlinear machines).*

**Remark II.2.** *The use of the* discontinuous *anti-windup decision function in* (10) *may lead to chattering [10]. If chattering occurs, the use of a Lipschitz continuous anti-windup decision function, as proposed in [6, 16], might be beneficial.*

### *D. Transfer function of the PI controller (without anti-windup)*

Using the notation $\boldsymbol{x}(t) \bullet\!\!-\!\!\circ\, \boldsymbol{x}(s)$ for the Laplace transform (assuming it exists) of some signal $\boldsymbol{x}(\cdot)$ and $\boldsymbol{x}(0+)$ for the right-handed initial value of $\boldsymbol{x}(\cdot)$, the Laplace transform of the PI controller (11) *without anti-windup* (i.e. neglecting the anti-windup decision function in (11)) is given by

$$\left.\begin{aligned} \tfrac{\mathrm{d}}{\mathrm{d}t}\boldsymbol{\xi}^{dq}_{\mathrm{i}}(t) &= \boldsymbol{e}^{dq}(t), \\ \boldsymbol{u}^{dq}_{\mathrm{pi}}(t) &= \boldsymbol{K}^{dq}_{\mathrm{p}}\, \boldsymbol{e}^{dq}(t) + \boldsymbol{K}^{dq}_{\mathrm{i}}\, \boldsymbol{\xi}^{dq}_{\mathrm{i}}(t) \end{aligned}\right\} \quad \circ\!\!-\!\!\bullet \quad \left\{\begin{aligned} s\boldsymbol{\xi}^{dq}_{\mathrm{i}}(s) + \boldsymbol{\xi}^{dq}_{\mathrm{i}}(0+) &= \boldsymbol{e}^{dq}(s), \\ \boldsymbol{u}^{dq}_{\mathrm{pi}}(s) &= \boldsymbol{K}^{dq}_{\mathrm{p}}\, \boldsymbol{e}^{dq}(s) + \boldsymbol{K}^{dq}_{\mathrm{i}}\, \boldsymbol{\xi}^{dq}_{\mathrm{i}}(s). \end{aligned}\right. \tag{13}$$

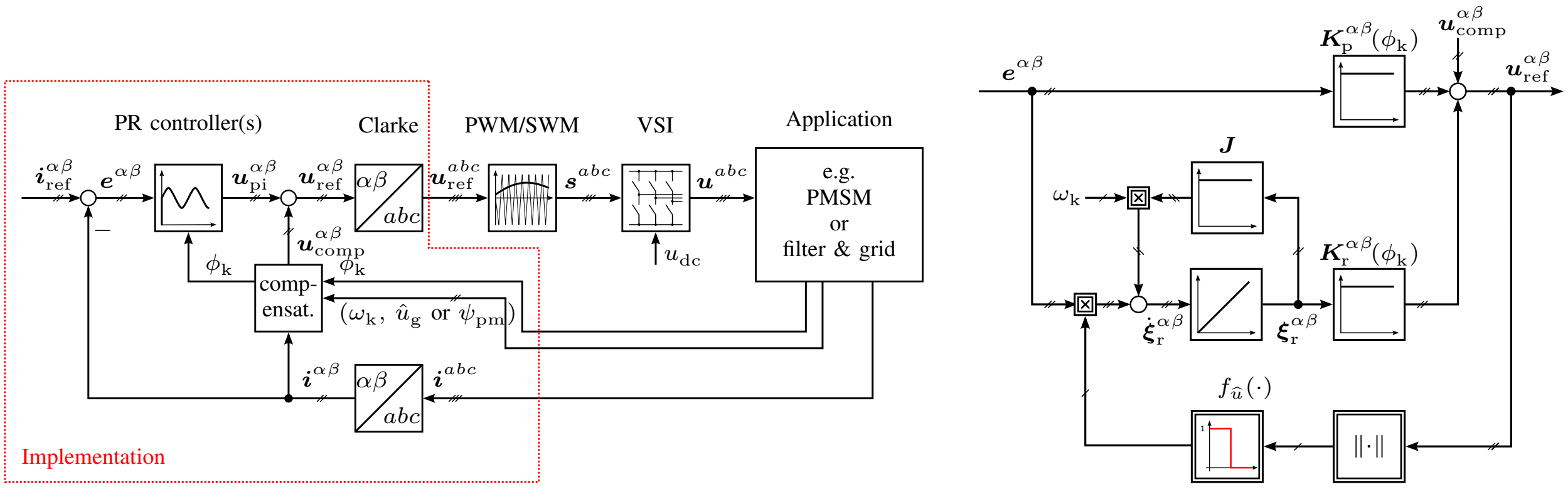

(a) *Overall controller structure.* (b) *PR controller with anti-windup.*

Figure 2: *Block diagrams of the overall controller structure and the PR controller with anti-windup in the stationary $(\alpha, \beta)$-reference frame and with disturbance feedforward compensation.*

Setting $\boldsymbol{\xi}_{\rm i}^{dq}(0+) = \mathbf{0}_2$, solving for $\boldsymbol{\xi}_{\rm i}^{dq}(s)$ and inserting into the last equation in (13) yields the well known transfer function

$$\boldsymbol{u}_{\rm pi}^{dq}(s) = \left[\boldsymbol{K}_{\rm p}^{dq} + \tfrac{1}{s}\boldsymbol{K}_{\rm i}^{dq}\right]\boldsymbol{e}^{dq}(s) = \begin{bmatrix} \frac{k_{\rm p}^{d}\, s + k_{\rm i}^{d}}{s} & 0 \\ 0 & \frac{k_{\rm p}^{q}\, s + k_{\rm i}^{q}}{s} \end{bmatrix} \boldsymbol{e}^{dq}(s) \tag{14}$$

of the PI controller(s) in the $(d, q)$-reference frame. Note that (14) *cannot* be implemented directly (it is a representation in the frequency domain). The state space representation (11) (including anti-windup) allows for a more general analysis (including nonlinear systems) and it is better suited for implementation (e.g. the differential equation can directly be discretized using Euler's method).

## III. Proportional-Resonant (PR) Controller with Anti-Windup in the $(\alpha, \beta)$-Reference Frame

In this section, a state space realization of proportional-resonant (PR) controller with anti-windup is proposed in the $(\alpha, \beta)$-reference frame.

### A. *Controller structure (see Fig. 2a)*

The proposed PR controller structure consists also of two parts, i.e.

$$\boldsymbol{u}_{\rm ref}^{\alpha\beta}(t) = \underbrace{\boldsymbol{u}_{\rm pr}^{\alpha\beta}(t)}_{\text{PR controller output}} + \underbrace{\boldsymbol{u}_{\rm comp}^{\alpha\beta}(t)}_{\text{disturbance compensation}}. \tag{15}$$

The voltage reference $\boldsymbol{u}_{\rm ref}^{\alpha\beta} = (u_{\rm ref}^{\alpha}, u_{\rm ref}^{\beta})^{\top}$ – the control input to the VSI – is the sum of the disturbance compensation $\boldsymbol{u}_{\rm comp}^{\alpha\beta} = (\boldsymbol{u}_{\rm comp}^{\alpha\beta}, \boldsymbol{u}_{\rm comp}^{\alpha\beta})^{\top}$ and the output $\boldsymbol{u}_{\rm pr}^{\alpha\beta} = (u_{\rm pr}^{\alpha}, u_{\rm pr}^{\beta})^{\top}$ of the PR controllers. Both parts will be discussed in more detail in the following.

### B. *Disturbance compensation (feedforward control)*

The dynamic models (4) and (6) in the $(\alpha, \beta)$-reference frame do not exhibit the coupling terms as the dynamic models (1) and (5) in the $(d, q)$-reference frame, respectively. Hence, only the following simpler disturbance terms

$$\boldsymbol{u}_{\rm dist}^{\alpha\beta}(t) := \begin{cases} -\boldsymbol{\psi}_{\rm pm}^{\alpha\beta}(t), & \text{for PMSMs as in (4)} \\ -\boldsymbol{u}_{\rm g}^{s}(t), & \text{for RL-filter \& grid as in (6)} \end{cases} \tag{16}$$

must be considered for the disturbance compensation design in the $(\alpha, \beta)$-reference frame. The disturbance terms (16) can be compensated for by the following feedforward control

$$\boldsymbol{u}_{\rm comp}^{\alpha\beta}(t) := \begin{pmatrix} u_{\rm comp}^{\alpha}(t) \\ u_{\rm comp}^{\beta}(t) \end{pmatrix} = -\boldsymbol{u}_{\rm dist}^{\alpha\beta}(t + T_{\rm delay})\boldsymbol{u}_{\rm comp}^{\alpha\beta}(s) = -\mathrm{e}^{sT_{\rm delay}}\boldsymbol{u}_{\rm dist}^{\alpha\beta}(s) \approx -\underbrace{\frac{c_0(1+sT_{\rm delay})}{1+sT_0}}_{=:F_{\rm comp}(s)}\boldsymbol{u}_{\rm dist}^{\alpha\beta}(s), \tag{17}$$

where $0 < c_0 \leq 1$ and $0 < T_0 \ll T_{\rm delay}$ are the same tuning parameters as introduced in Sec. II-B.

### C. PR controller with anti-windup (see Fig. 2b)

Similarly to the PI controller, a simple but effective anti-windup strategy is utilized to stop integration of the integral control action of the PR controller (here: $\boldsymbol{u}^{\alpha\beta}$ or $\boldsymbol{u}^{\alpha\beta}_{\mathrm{ref}}$) if it exceeds its admissible range. For this, the "anti-windup decision function" as in (10) (with the argument $\boldsymbol{u}^{\alpha\beta}_{\mathrm{ref}}$ instead of $\boldsymbol{u}^{dq}_{\mathrm{ref}}$) is merged with the PR controller as follows

$$\left.\boxed{\begin{aligned} \tfrac{\mathrm{d}}{\mathrm{d}t}\boldsymbol{\xi}^{\alpha\beta}_{\mathrm{r}}(t) &= f_{\widehat{u}}\big(\|\boldsymbol{u}^{\alpha\beta}_{\mathrm{ref}}(t)\|\big)\,\boldsymbol{e}^{\alpha\beta}(t) + \omega_{\mathrm{k}}(t)\boldsymbol{J}\boldsymbol{\xi}^{\alpha\beta}_{\mathrm{r}}(t), \qquad \boldsymbol{\xi}^{\alpha\beta}_{\mathrm{r}}(0) = \boldsymbol{\xi}^{\alpha\beta,0}_{\mathrm{r}} \in \mathbb{R}^2 \\ \boldsymbol{u}^{\alpha\beta}_{\mathrm{pr}}(t) &= \boldsymbol{K}^{\alpha\beta}_{\mathrm{p}}(\phi_{\mathrm{k}}(t))\,\boldsymbol{e}^{\alpha\beta}(t) + \boldsymbol{K}^{\alpha\beta}_{\mathrm{r}}(\phi_{\mathrm{k}}(t))\,\boldsymbol{\xi}^{\alpha\beta}_{\mathrm{r}}(t), \end{aligned}}\right\} \tag{18}$$

where $\boldsymbol{\xi}^{\alpha\beta}_{\mathrm{r}} = (\xi^{\alpha}_{\mathrm{r}}, \xi^{\beta}_{\mathrm{r}})^{\top}$ is the output vector of the internal state of the PR controller, $\boldsymbol{\xi}^{\alpha\beta,0}_{\mathrm{r}}$ is its initial value vector and $\boldsymbol{e}^{\alpha\beta} = (e^{\alpha}, e^{\beta})^{\top}$ is again the tracking error vector but now in the $(\alpha,\beta)$ reference frame. The controller gains are merged into the gain matrices

$$\boldsymbol{K}^{\alpha\beta}_{\mathrm{p}}(\phi_{\mathrm{k}}) := \boldsymbol{T}_{\mathrm{p}}(\phi_{\mathrm{k}})\begin{bmatrix} k^{\alpha}_{\mathrm{p}} & 0 \\ 0 & k^{\beta}_{\mathrm{p}}\end{bmatrix}\boldsymbol{T}_{\mathrm{p}}(\phi_{\mathrm{k}})^{-1} = \begin{bmatrix} k^{\alpha}_{\mathrm{p}} - (k^{\alpha}_{\mathrm{p}} - k^{\beta}_{\mathrm{p}})\sin(\phi_{\mathrm{k}})^2 & \frac{1}{2}\sin(2\phi_{\mathrm{k}})(k^{\alpha}_{\mathrm{p}} - k^{\beta}_{\mathrm{p}}) \\ \frac{1}{2}\sin(2\phi_{\mathrm{k}})(k^{\alpha}_{\mathrm{p}} - k^{\beta}_{\mathrm{p}}) & k^{\beta}_{\mathrm{p}} + (k^{\alpha}_{\mathrm{p}} - k^{\beta}_{\mathrm{p}})\sin(\phi_{\mathrm{k}})^2 \end{bmatrix} \quad \text{and}$$

$$\boldsymbol{K}^{\alpha\beta}_{\mathrm{r}}(\phi_{\mathrm{k}}) := \boldsymbol{T}_{\mathrm{p}}(\phi_{\mathrm{k}})\begin{bmatrix} k^{\alpha}_{\mathrm{r}} & 0 \\ 0 & k^{\beta}_{\mathrm{r}}\end{bmatrix}\boldsymbol{T}_{\mathrm{p}}(\phi_{\mathrm{k}})^{-1} = \begin{bmatrix} k^{\alpha}_{\mathrm{r}} - (k^{\alpha}_{\mathrm{r}} - k^{\beta}_{\mathrm{r}})\sin(\phi_{\mathrm{k}})^2 & \frac{1}{2}\sin(2\phi_{\mathrm{k}})(k^{\alpha}_{\mathrm{r}} - k^{\beta}_{\mathrm{r}}) \\ \frac{1}{2}\sin(2\phi_{\mathrm{k}})(k^{\alpha}_{\mathrm{r}} - k^{\beta}_{\mathrm{r}}) & k^{\beta}_{\mathrm{r}} + (k^{\alpha}_{\mathrm{r}} - k^{\beta}_{\mathrm{r}})\sin(\phi_{\mathrm{k}})^2 \end{bmatrix}, \tag{19}$$

which, in the most general case, *depend* on the electrical angle $\phi_{\mathrm{k}}$ (used for the Park transformation $\boldsymbol{T}_{\mathrm{p}}(\phi_{\mathrm{k}})$) and are *not* diagonal. The gain matrices $\boldsymbol{K}^{\alpha\beta}_{\mathrm{p}}(\phi_{\mathrm{k}})$ and $\boldsymbol{K}^{\alpha\beta}_{\mathrm{r}}(\phi_{\mathrm{k}})$ are diagonal if and only if (i) $k^{\alpha}_{\mathrm{p}} = k^{\beta}_{\mathrm{p}}$ and $k^{\alpha}_{\mathrm{r}} = k^{\beta}_{\mathrm{r}}$ or (ii) $\phi_{\mathrm{k}} = l\pi$, $l \in \mathbb{N}$ (which is only mathematically of interest; for machine-side or grid-side control, $\phi_{\mathrm{k}}$ will change with the mechanical angular frequency or the grid angular frequency). Finally, note that the anti-windup decision function $f_{\widehat{u}}\big(\|\boldsymbol{u}^{\alpha\beta}_{\mathrm{ref}}\|\big)$ in (18) will disable the effect of the tracking error $\boldsymbol{e}^{\alpha\beta}$ on the derivative $\dot{\boldsymbol{\xi}}^{\alpha\beta}_{\mathrm{r}}$ of the PR integrator states whereas the capability of reduplicating sinusoidal signals by $\boldsymbol{\xi}^{\alpha\beta}_{\mathrm{r}}$ is preserved.

**Remark III.1.** *Note that, for example, for anisotropic PMSMs or reluctance synchronous machines (RSMs) with $L^{d}_{\mathrm{s}} \neq L^{q}_{\mathrm{s}}$, the proportional gains $k^{\alpha}_{\mathrm{p}}$ and $k^{\beta}_{\mathrm{p}}$ should be chosen differently [14, 15].*

### D. Transfer function of the PR controller (without anti-windup and with constant angular velocity)

To derive the transfer functions of the PR controllers, the following assumption must be imposed:

**Assumption III.2.**

- *The anti-windup decision function in* (18) *is neglected, i.e.* $f_{\widehat{u}}\big(\|\boldsymbol{u}^{\alpha\beta}_{\mathrm{ref}}\|\big) = 1$ *for all* $\boldsymbol{u}^{\alpha\beta}_{\mathrm{ref}} \in \mathbb{R}^2$*;*
- *The angular frequency is constant (and positive), i.e.* $\omega_{\mathrm{k}}(t) = \omega_{\mathrm{k}} > 0$ *for all* $t \geq 0$*; and*
- *The proportional and resonant controller gains are chosen to be equal (and positive), respectively, i.e.*

$$k^{\alpha}_{\mathrm{p}} = k^{\beta}_{\mathrm{p}} =: k_{\mathrm{p}} > 0 \qquad \textit{and} \qquad k^{\alpha}_{\mathrm{r}} = k^{\beta}_{\mathrm{r}} =: k_{\mathrm{r}} > 0.$$

Note that, without imposing Assumption III.2, the transfer function could not be derived. The Laplace transform of the PR controllers (18) (if and only if Assumption III.2 holds) is given by

$$\left.\begin{aligned} \tfrac{\mathrm{d}}{\mathrm{d}t}\boldsymbol{\xi}^{\alpha\beta}_{\mathrm{r}}(t) &= \boldsymbol{e}^{\alpha\beta}(t) + \omega_{\mathrm{k}}\boldsymbol{J}\boldsymbol{\xi}^{\alpha\beta}_{\mathrm{r}}(t), \\ \boldsymbol{u}^{\alpha\beta}_{\mathrm{pr}}(t) &= k_{\mathrm{p}}\boldsymbol{e}^{\alpha\beta}(t) + k_{\mathrm{r}}\boldsymbol{\xi}^{\alpha\beta}_{\mathrm{r}}(t) \end{aligned}\right\} \quad \circ\!\!-\!\!\bullet \quad \left\{\begin{aligned} s\boldsymbol{\xi}^{\alpha\beta}_{\mathrm{r}}(s) + \boldsymbol{\xi}^{\alpha\beta}_{\mathrm{r}}(0+) &= \boldsymbol{e}^{\alpha\beta}(s) + \omega_{\mathrm{k}}\boldsymbol{J}\boldsymbol{\xi}^{\alpha\beta}_{\mathrm{r}}(s), \\ \boldsymbol{u}^{\alpha\beta}_{\mathrm{pr}}(s) &= k_{\mathrm{p}}\,\boldsymbol{e}^{\alpha\beta}(s) + k_{\mathrm{r}}\,\boldsymbol{\xi}^{\alpha\beta}_{\mathrm{r}}(s). \end{aligned}\right. \tag{20}$$

Setting $\boldsymbol{\xi}^{\alpha\beta}_{\mathrm{r}}(0+) = \boldsymbol{0}_2$ and solving for $\boldsymbol{\xi}^{\alpha\beta}_{\mathrm{r}}(s)$ in the first row of equation (20) gives

$$\boldsymbol{\xi}^{\alpha\beta}_{\mathrm{r}}(s)\Big[s\boldsymbol{I}_2 - \omega_{\mathrm{k}}\boldsymbol{J}\Big] = \boldsymbol{\xi}^{\alpha\beta}_{\mathrm{r}}(s)\underbrace{\begin{bmatrix} s & \omega_{\mathrm{k}} \\ -\omega_{\mathrm{k}} & s\end{bmatrix}}_{=:\boldsymbol{P}(s)} = \boldsymbol{e}^{\alpha\beta}(s) \qquad \Longrightarrow \qquad \boldsymbol{\xi}^{\alpha\beta}_{\mathrm{r}}(s) = \underbrace{\tfrac{1}{s^2+\omega_{\mathrm{k}}^2}\begin{bmatrix} s & -\omega_{\mathrm{k}} \\ \omega_{\mathrm{k}} & s\end{bmatrix}}_{=:\boldsymbol{P}(s)^{-1}}\boldsymbol{e}^{\alpha\beta}(s). \tag{21}$$

Inserting (21) into the right-hand side of equation (20) yields the well known transfer function

$$\boldsymbol{u}^{\alpha\beta}_{\mathrm{pr}}(s) = \Big[k_{\mathrm{p}}\boldsymbol{I}_2 + k_{\mathrm{r}}\boldsymbol{P}(s)^{-1}\Big]\boldsymbol{e}^{\alpha\beta}(s) = \begin{bmatrix} k_{\mathrm{p}} + k_{\mathrm{r}}\frac{s}{s^2+\omega_{\mathrm{k}}^2}, & -k_{\mathrm{r}}\frac{\omega_{\mathrm{k}}}{s^2+\omega_{\mathrm{k}}^2} \\ k_{\mathrm{r}}\frac{\omega_{\mathrm{k}}}{s^2+\omega_{\mathrm{k}}^2}, & k_{\mathrm{p}} + k_{\mathrm{r}}\frac{s}{s^2+\omega_{\mathrm{k}}^2}\end{bmatrix}\boldsymbol{e}^{\alpha\beta}(s) \tag{22}$$

of the PR controller(s) in the $(\alpha,\beta)$-reference frame. Note that the off-diagonal terms in (22) are (usually) not considered and represent actually a cross-coupling of the PR controller. This cross-coupling cancels out if positive and negative zero sequence are considered [1, Appendix C].

**Remark III.3.** *Due to the infinite gain of the transfer function* $\frac{s}{s^2+\omega_k^2}$ *at* $s = \pm j\omega_k$*, an implementation of the following approximation* $\frac{\omega_c\, s}{s^2+2\omega_c\, s+\omega_k^2} \approx \frac{s}{s^2+\omega_k^2}$ *is often recommended [1, C. 3] which gives the following approximation*

$$\widehat{\boldsymbol{u}}_{\mathrm{pr}}^{\alpha\beta}(s) = \begin{bmatrix} k_{\mathrm{p}} + k_{\mathrm{r}} \frac{\omega_c\, s}{s^2+2\,\omega_c\, s+\omega_k^2}, & -k_{\mathrm{r}} \frac{\omega_c\, \omega_k}{s^2+2\,\omega_c\, s+\omega_k^2} \\ k_{\mathrm{r}} \frac{\omega_c\, \omega_k}{s^2+2\,\omega_c\, s+\omega_k^2}, & k_{\mathrm{p}} + k_{\mathrm{r}} \frac{\omega_c\, s}{s^2+2\,\omega_c\, s+\omega_k^2} \end{bmatrix} \boldsymbol{e}^{dq}(s) \approx \boldsymbol{u}_{\mathrm{pr}}^{\alpha\beta}(s) \tag{23}$$

*of the PR controller in the frequency domain. Note that, the state space implementation* (18) *does not require such an approximation. Moreover, the implementation of the approximated PR controller as in* (23) *will exhibit a different transient behavior than that of the PR controller* (18).

## IV. Equivalence of PI and PR controller with anti-windup in state space

In this section, the main result is presented: PI controller (11) and PR controller (18) are equivalent in the synchronously rotating $(d, q)$ reference frame *and* in the stationary $(\alpha, \beta)$ reference frame if and only if the controller parameters and the initial values are chosen identically.

### *A. Equivalence of PI and PR controller with anti-windup in the synchronously rotating* $(d, q)$*-reference frame*

First note that $\frac{\mathrm{d}}{\mathrm{d}t}\phi_k(t) = \omega_k(t)$ and $\frac{\mathrm{d}}{\mathrm{d}t}\boldsymbol{T}_{\mathrm{p}}(\phi_k(t)) = \omega_k(t)\boldsymbol{J}\boldsymbol{T}_{\mathrm{p}}(\phi_k(t))$ for all $t \geq 0$ (if $\phi_k(\cdot) \in \mathcal{C}^1(\mathbb{R}_{\geq 0}; \mathbb{R})$). Moreover, $\boldsymbol{J}\boldsymbol{T}_{\mathrm{p}}(\phi_k) = \boldsymbol{T}_{\mathrm{p}}(\phi_k)\boldsymbol{J}$ (matrices commute) and $\boldsymbol{T}_{\mathrm{p}}(\phi_k)\boldsymbol{T}_{\mathrm{p}}(\phi_k)^{-1} = \boldsymbol{T}_{\mathrm{p}}(\phi_k)^{-1}\boldsymbol{T}_{\mathrm{p}}(\phi_k) = \boldsymbol{I}_2$ hold for all $\phi_k \in \mathbb{R}$. Then, applying the Park transformation (with $\boldsymbol{x}^{dq} = \boldsymbol{T}_{\mathrm{p}}(\phi_k)^{-1}\boldsymbol{x}^{\alpha\beta}$) and the product rule of differentiation to the *left-* and *right*-hand side of the first row in (18) yield

$$\begin{aligned}\boldsymbol{T}_{\mathrm{p}}(\phi_k(t))^{-1}\tfrac{\mathrm{d}}{\mathrm{d}t}\boldsymbol{\xi}_{\mathrm{r}}^{\alpha\beta}(t) &= \boldsymbol{T}_{\mathrm{p}}(\phi_k(t))^{-1}\Big(\tfrac{\mathrm{d}}{\mathrm{d}t}\underbrace{\boldsymbol{T}_{\mathrm{p}}(\phi_k(t))\boldsymbol{\xi}_{\mathrm{r}}^{dq}(t)}_{=\boldsymbol{\xi}_{\mathrm{r}}^{\alpha\beta}(t)}\Big) = \boldsymbol{T}_{\mathrm{p}}(\phi_k(t))^{-1}\Big(\big(\tfrac{\mathrm{d}}{\mathrm{d}t}\boldsymbol{T}_{\mathrm{p}}(\phi_k(t))\big)\boldsymbol{\xi}_{\mathrm{r}}^{dq}(t) + \boldsymbol{T}_{\mathrm{p}}(\phi_k(t))\tfrac{\mathrm{d}}{\mathrm{d}t}\boldsymbol{\xi}_{\mathrm{r}}^{dq}(t)\Big) \\ &= \boldsymbol{T}_{\mathrm{p}}(\phi_k(t))^{-1}\omega_k(t)\boldsymbol{J}\boldsymbol{T}_{\mathrm{p}}(\phi_k(t))\boldsymbol{\xi}_{\mathrm{r}}^{dq}(t) + \boldsymbol{T}_{\mathrm{p}}(\phi_k(t))^{-1}\boldsymbol{T}_{\mathrm{p}}(\phi_k(t))\tfrac{\mathrm{d}}{\mathrm{d}t}\boldsymbol{\xi}_{\mathrm{r}}^{dq}(t) \\ &= \omega_k(t)\boldsymbol{J}\boldsymbol{\xi}_{\mathrm{r}}^{dq}(t) + \tfrac{\mathrm{d}}{\mathrm{d}t}\boldsymbol{\xi}_{\mathrm{r}}^{dq}(t)\end{aligned} \tag{24}$$

and

$$\begin{aligned}\boldsymbol{T}_{\mathrm{p}}(\phi_k(t))^{-1}\tfrac{\mathrm{d}}{\mathrm{d}t}\boldsymbol{\xi}_{\mathrm{r}}^{\alpha\beta}(t) &\overset{(18)}{=} \boldsymbol{T}_{\mathrm{p}}(\phi_k(t))^{-1}\Big(f_{\widehat{u}}\big(\|\underbrace{\boldsymbol{T}_{\mathrm{p}}(\phi_k(t))\boldsymbol{u}_{\mathrm{ref}}^{dq}(t)}_{=\boldsymbol{u}_{\mathrm{ref}}^{\alpha\beta}(t)}\|\big)\underbrace{\boldsymbol{T}_{\mathrm{p}}(\phi_k(t))\boldsymbol{e}^{dq}(t)}_{=\boldsymbol{e}^{\alpha\beta}(t)} + \omega_k(t)\boldsymbol{J}\underbrace{\boldsymbol{T}_{\mathrm{p}}(\phi_k(t))\boldsymbol{\xi}_{\mathrm{r}}^{dq}(t)}_{=\boldsymbol{\xi}_{\mathrm{r}}^{\alpha\beta}(t)}\Big) \\ &= f_{\widehat{u}}\big(\|\boldsymbol{u}_{\mathrm{ref}}^{dq}(t)\|\big)\boldsymbol{e}^{dq}(t) + \omega_k(t)\boldsymbol{J}\boldsymbol{\xi}_{\mathrm{r}}^{dq}(t),\end{aligned} \tag{25}$$

respectively. Combining these two equations gives

$$\begin{aligned}\boldsymbol{T}_{\mathrm{p}}(\phi_k(t))^{-1}\tfrac{\mathrm{d}}{\mathrm{d}t}\boldsymbol{\xi}_{\mathrm{r}}^{\alpha\beta}(t) \overset{(24)}{=} \cancel{\omega_k(t)\boldsymbol{J}\boldsymbol{\xi}_{\mathrm{r}}^{dq}(t)} + \tfrac{\mathrm{d}}{\mathrm{d}t}\boldsymbol{\xi}_{\mathrm{r}}^{dq}(t) &\overset{(25)}{=} f_{\widehat{u}}\big(\|\boldsymbol{u}_{\mathrm{ref}}^{dq}(t)\|\big)\boldsymbol{e}^{dq}(t) + \cancel{\omega_k(t)\boldsymbol{J}\boldsymbol{\xi}_{\mathrm{r}}^{dq}(t)} \\ &\implies \tfrac{\mathrm{d}}{\mathrm{d}t}\boldsymbol{\xi}_{\mathrm{r}}^{dq}(t) = f_{\widehat{u}}\big(\|\boldsymbol{u}_{\mathrm{ref}}^{dq}(t)\|\big)\boldsymbol{e}^{dq}(t),\end{aligned} \tag{26}$$

which is *similar* to the first equation in (11) except for possibly differing initial values $\boldsymbol{\xi}_{\mathrm{r}}^{dq}(0)$ and $\boldsymbol{\xi}_{\mathrm{i}}^{dq}(0)$. Now, applying the Park transformation to the second row in (18) and considering the definition of the gain matrices in (19) leads to

$$\begin{aligned}\boldsymbol{u}_{\mathrm{pr}}^{dq}(t) &:= \boldsymbol{T}_{\mathrm{p}}(\phi_k(t))^{-1}\boldsymbol{u}_{\mathrm{pr}}^{\alpha\beta}(t) \overset{(18)}{=} \boldsymbol{T}_{\mathrm{p}}(\phi_k(t))^{-1}\Big(\boldsymbol{K}_{\mathrm{p}}^{\alpha\beta}(\phi_k(t))\underbrace{\boldsymbol{T}_{\mathrm{p}}(\phi_k(t))\boldsymbol{e}^{dq}(t)}_{=\boldsymbol{e}^{\alpha\beta}(t)} + \boldsymbol{K}_{\mathrm{r}}^{\alpha\beta}(\phi_k(t))\underbrace{\boldsymbol{T}_{\mathrm{p}}(\phi_k(t))\boldsymbol{\xi}_{\mathrm{r}}^{dq}(t)}_{=\boldsymbol{\xi}_{\mathrm{r}}^{\alpha\beta}(t)}\Big) \\ &\overset{(19)}{=} \boldsymbol{T}_{\mathrm{p}}(\phi_k(t))^{-1}\Big(\boldsymbol{T}_{\mathrm{p}}(\phi_k(t))\begin{bmatrix} k_{\mathrm{p}}^{\alpha} & 0 \\ 0 & k_{\mathrm{p}}^{\beta}\end{bmatrix}\boldsymbol{T}_{\mathrm{p}}(\phi_k(t))^{-1}\boldsymbol{T}_{\mathrm{p}}(\phi_k(t))\boldsymbol{e}^{dq}(t) + \boldsymbol{T}_{\mathrm{p}}(\phi_k(t))\begin{bmatrix} k_{\mathrm{r}}^{\alpha} & 0 \\ 0 & k_{\mathrm{r}}^{\beta}\end{bmatrix}\boldsymbol{T}_{\mathrm{p}}(\phi_k(t))^{-1}\boldsymbol{T}_{\mathrm{p}}(\phi_k(t))\boldsymbol{\xi}_{\mathrm{r}}^{dq}(t)\Big) \\ &= \begin{bmatrix} k_{\mathrm{p}}^{\alpha} & 0 \\ 0 & k_{\mathrm{p}}^{\beta}\end{bmatrix}\boldsymbol{e}^{dq}(t) + \begin{bmatrix} k_{\mathrm{r}}^{\alpha} & 0 \\ 0 & k_{\mathrm{r}}^{\beta}\end{bmatrix}\boldsymbol{\xi}_{\mathrm{r}}^{dq}(t),\end{aligned} \tag{27}$$

which is *similar* to the second equation in (11). Now, by setting

- $\boldsymbol{\xi}_{\mathrm{r}}^{dq} \overset{!}{=} \boldsymbol{\xi}_{\mathrm{i}}^{dq}$ with $\boldsymbol{\xi}_{\mathrm{r}}^{dq}(0) = \boldsymbol{T}_{\mathrm{p}}(\phi_k(0))^{-1}\boldsymbol{\xi}_{\mathrm{r}}^{\alpha\beta,0} \overset{!}{=} \boldsymbol{\xi}_{\mathrm{i}}^{dq,0}$ (where $\boldsymbol{\xi}_{\mathrm{i}}^{dq,0}$ is as in (11)), and
- $k_{\mathrm{p}}^{\alpha} \overset{!}{=} k_{\mathrm{p}}^{d}$, $k_{\mathrm{p}}^{\beta} \overset{!}{=} k_{\mathrm{p}}^{q}$, $k_{\mathrm{r}}^{\alpha} \overset{!}{=} k_{\mathrm{i}}^{d}$ and $k_{\mathrm{r}}^{\beta} \overset{!}{=} k_{\mathrm{i}}^{q}$,

the PR controller (18) and the PI controller (11) are *identical*, i.e. $\boldsymbol{u}_{\mathrm{pi}}^{dq}(t) = \boldsymbol{u}_{\mathrm{pr}}^{dq}(t)$ holds for all $t \geq 0$ in the $(d, q)$-reference frame. Hence, PR and PI controller are *equivalent* in the $(d, q)$-reference frame.

*B. Equivalence of PI and PR controller with anti-windup in the stationary $(\alpha,\beta)$-reference frame*

Now, the reverse is shown. Applying the (inverse) Park transformation (with $\boldsymbol{x}^{\alpha\beta} = \boldsymbol{T}_{\mathrm{p}}(\phi_{\mathrm{k}})\boldsymbol{x}^{dq}$) and the product rule of differentiation to the *left-* and *right-*hand side of the first row in (11) yield

$$\begin{aligned}\boldsymbol{T}_{\mathrm{p}}(\phi_{\mathrm{k}}(t))\tfrac{\mathrm{d}}{\mathrm{d}t}\boldsymbol{\xi}_{\mathrm{i}}^{dq}(t) &= \boldsymbol{T}_{\mathrm{p}}(\phi_{\mathrm{k}}(t))\Big(\tfrac{\mathrm{d}}{\mathrm{d}t}\underbrace{\boldsymbol{T}_{\mathrm{p}}(\phi_{\mathrm{k}}(t))^{-1}\boldsymbol{\xi}_{\mathrm{i}}^{\alpha\beta}(t)}_{=\boldsymbol{\xi}_{\mathrm{i}}^{dq}(t)}\Big) = \boldsymbol{T}_{\mathrm{p}}(\phi_{\mathrm{k}}(t))\Big(\big(\tfrac{\mathrm{d}}{\mathrm{d}t}\boldsymbol{T}_{\mathrm{p}}(\phi_{\mathrm{k}}(t))^{-1}\big)\boldsymbol{\xi}_{\mathrm{i}}^{\alpha\beta}(t) + \boldsymbol{T}_{\mathrm{p}}(\phi_{\mathrm{k}}(t))^{-1}\tfrac{\mathrm{d}}{\mathrm{d}t}\boldsymbol{\xi}_{\mathrm{i}}^{\alpha\beta}(t)\Big) \\ &= -\boldsymbol{T}_{\mathrm{p}}(\phi_{\mathrm{k}}(t))\omega_{\mathrm{k}}(t)\boldsymbol{J}\boldsymbol{T}_{\mathrm{p}}(\phi_{\mathrm{k}}(t))^{-1}\boldsymbol{\xi}_{\mathrm{i}}^{\alpha\beta}(t) + \boldsymbol{T}_{\mathrm{p}}(\phi_{\mathrm{k}}(t))\boldsymbol{T}_{\mathrm{p}}(\phi_{\mathrm{k}}(t))^{-1}\tfrac{\mathrm{d}}{\mathrm{d}t}\boldsymbol{\xi}_{\mathrm{i}}^{\alpha\beta}(t) \\ &= -\omega_{\mathrm{k}}(t)\boldsymbol{J}\boldsymbol{\xi}_{\mathrm{i}}^{\alpha\beta}(t) + \tfrac{\mathrm{d}}{\mathrm{d}t}\boldsymbol{\xi}_{\mathrm{i}}^{\alpha\beta}(t) \end{aligned} \tag{28}$$

and

$$\boldsymbol{T}_{\mathrm{p}}(\phi_{\mathrm{k}}(t))\tfrac{\mathrm{d}}{\mathrm{d}t}\boldsymbol{\xi}_{\mathrm{i}}^{dq}(t) \overset{(11)}{=} \boldsymbol{T}_{\mathrm{p}}(\phi_{\mathrm{k}}(t))\Big(f_{\widehat{u}}\big(\|\underbrace{\boldsymbol{T}_{\mathrm{p}}(\phi_{\mathrm{k}}(t))^{-1}\boldsymbol{u}_{\mathrm{ref}}^{\alpha\beta}(t)}_{=\boldsymbol{u}_{\mathrm{ref}}^{dq}(t)}\|\big)\underbrace{\boldsymbol{T}_{\mathrm{p}}(\phi_{\mathrm{k}}(t))^{-1}\boldsymbol{e}^{\alpha\beta}(t)}_{=\boldsymbol{e}^{dq}(t)}\Big) = f_{\widehat{u}}\big(\boldsymbol{u}_{\mathrm{ref}}^{\alpha\beta}(t)\big)\boldsymbol{e}^{\alpha\beta}(t), \tag{29}$$

respectively. Setting these two equations equal leads to

$$\begin{aligned}\boldsymbol{T}_{\mathrm{p}}(\phi_{\mathrm{k}}(t))\tfrac{\mathrm{d}}{\mathrm{d}t}\boldsymbol{\xi}_{\mathrm{i}}^{dq}(t) &\overset{(28)}{=} -\omega_{\mathrm{k}}(t)\boldsymbol{J}\boldsymbol{\xi}_{\mathrm{i}}^{\alpha\beta}(t) + \tfrac{\mathrm{d}}{\mathrm{d}t}\boldsymbol{\xi}_{\mathrm{i}}^{\alpha\beta}(t) \overset{(29)}{=} f_{\widehat{u}}\big(\|\boldsymbol{u}_{\mathrm{ref}}^{\alpha\beta}(t)\|\big)\boldsymbol{e}^{\alpha\beta}(t) \\ &\implies \tfrac{\mathrm{d}}{\mathrm{d}t}\boldsymbol{\xi}_{\mathrm{i}}^{\alpha\beta}(t) = f_{\widehat{u}}\big(\|\boldsymbol{u}_{\mathrm{ref}}^{\alpha\beta}(t)\|\big)\boldsymbol{e}^{\alpha\beta}(t) + \omega_{\mathrm{k}}(t)\boldsymbol{J}\boldsymbol{\xi}_{\mathrm{i}}^{\alpha\beta}(t), \end{aligned} \tag{30}$$

which is *similar* to the first equation in (18) except for possibly differing initial values $\boldsymbol{\xi}_{\mathrm{i}}^{\alpha\beta}(0)$ and $\boldsymbol{\xi}_{\mathrm{r}}^{\alpha\beta}(0)$. Now, applying the Park transformation to the second row in (11) and considering the definition of the gain matrices in (12) leads to

$$\begin{aligned}\boldsymbol{u}_{\mathrm{pi}}^{\alpha\beta}(t) := \boldsymbol{T}_{\mathrm{p}}(\phi_{\mathrm{k}}(t))\boldsymbol{u}_{\mathrm{pi}}^{dq}(t) &\overset{(11)}{=} \boldsymbol{T}_{\mathrm{p}}(\phi_{\mathrm{k}}(t))\Big(\boldsymbol{K}_{\mathrm{p}}^{dq}\underbrace{\boldsymbol{T}_{\mathrm{p}}(\phi_{\mathrm{k}}(t))^{-1}\boldsymbol{e}^{\alpha\beta}(t)}_{=\boldsymbol{e}^{dq}(t)} + \boldsymbol{K}_{\mathrm{i}}^{dq}\underbrace{\boldsymbol{T}_{\mathrm{p}}(\phi_{\mathrm{k}}(t))^{-1}\boldsymbol{\xi}_{\mathrm{i}}^{\alpha\beta}(t)}_{=\boldsymbol{\xi}_{\mathrm{i}}^{dq}(t)}\Big) \\ &= \underbrace{\boldsymbol{T}_{\mathrm{p}}(\phi_{\mathrm{k}}(t))\,\boldsymbol{K}_{\mathrm{p}}^{dq}\,\boldsymbol{T}_{\mathrm{p}}(\phi_{\mathrm{k}}(t))^{-1}}_{=:\boldsymbol{K}_{\mathrm{p}}^{\alpha\beta}(\phi_{\mathrm{k}})}\boldsymbol{e}^{\alpha\beta}(t) + \underbrace{\boldsymbol{T}_{\mathrm{p}}(\phi_{\mathrm{k}}(t))\,\boldsymbol{K}_{\mathrm{i}}^{dq}\,\boldsymbol{T}_{\mathrm{p}}(\phi_{\mathrm{k}}(t))^{-1}}_{=:\boldsymbol{K}_{\mathrm{i}}^{\alpha\beta}(\phi_{\mathrm{k}})}\boldsymbol{\xi}_{\mathrm{i}}^{\alpha\beta}(t), \end{aligned} \tag{31}$$

which is *similar* to the second equation in (18) taking the gain matrix definitions in (19) into account. Now, by setting

- $\boldsymbol{\xi}_{\mathrm{i}}^{\alpha\beta} \overset{!}{=} \boldsymbol{\xi}_{\mathrm{r}}^{\alpha\beta}$ with $\boldsymbol{\xi}_{\mathrm{i}}^{\alpha\beta}(0) = \boldsymbol{T}_{\mathrm{p}}(\phi_{\mathrm{k}}(0))\boldsymbol{\xi}_{\mathrm{i}}^{dq,0} \overset{!}{=} \boldsymbol{\xi}_{\mathrm{r}}^{\alpha\beta,0}$ (where $\boldsymbol{\xi}_{\mathrm{r}}^{\alpha\beta,0}$ is as in (18)), and
- $k_{\mathrm{p}}^{d} \overset{!}{=} k_{\mathrm{p}}^{\alpha}$, $k_{\mathrm{p}}^{q} \overset{!}{=} k_{\mathrm{p}}^{\beta}$, $k_{\mathrm{i}}^{d} \overset{!}{=} k_{\mathrm{r}}^{\alpha}$ and $k_{\mathrm{i}}^{q} \overset{!}{=} k_{\mathrm{r}}^{\beta}$,

the PI controller (11) and the PR controller (18) are *identical*, i.e. $\boldsymbol{u}_{\mathrm{pi}}^{dq}(t) = \boldsymbol{u}_{\mathrm{pr}}^{dq}(t)$ holds for all $t \geq 0$ in the $(\alpha,\beta)$-reference frame. Hence, PI and PR controller are also *equivalent* in the $(\alpha,\beta)$-reference frame.